\documentclass[aps,pra,twocolumn,showpacs]{revtex4}
\usepackage[dvips]{graphicx}
\usepackage{amsmath}
\usepackage{times}

\usepackage{color}
\usepackage{psfrag}
\usepackage{epic}
\usepackage{amssymb}
\usepackage{pst-all}

\begin{document}

\title{Ultracold bosons in lattices with binary disorder}

\author{K.~V.~Krutitsky,$^1$ M.~Thorwart,$^2$ R.~Egger,$^2$ and R.~Graham$^1$}
\affiliation{
$^1$Fachbereich Physik der Universit\"at Duisburg-Essen, Campus Duisburg,
    Lotharstr.~1, 47048 Duisburg, Germany\\
$^2$Institut f\"ur Theoretische Physik, Heinrich-Heine-Universit\"at,
    D-40225 D\"usseldorf, Germany
}

\date{\today}

\begin{abstract}
Quantum phases of ultracold bosons with repulsive interactions in lattices
in the presence of quenched disorder are investigated.
The disorder is assumed to be caused
by the interaction of the bosons with impurity atoms having a large effective mass.
The system is described by the Bose-Hubbard Hamiltonian
with random on-site energies which have a discrete binary probability distribution.
The phase diagram at zero temperature is calculated using several methods like
a strong-coupling expansion, an exact numerical diagonalization, and a
Bose-Fermi mapping valid in the hard-core limit.
It is shown that the Mott-insulator phase exists for any strength
of disorder in contrast to the case of continuous probability distribution.
We find  that the compressibility of the Bose glass phase varies in a wide range
and can be extremely low.
Furthermore, we evaluate
experimentally accessible quantities like the momentum distribution,
the static and dynamic structure factors, and the density of excited states.
The influence of finite temperature is discussed as well.
\end{abstract}


\maketitle

\section{Introduction}

The remarkable experimental control over ultracold atomic gases in optical lattices
acquired in recent years~\cite{Zwerger,Zoller,Bloch2,MO,review,BDZ} 
has opened up completely new lines of investigation in the field of strongly correlated quantum systems.
One of these are quantum phase transitions (QPTs) of ultracold
atoms in optical lattices.
These fascinating phenomena are caused by the interplay of quantum tunneling,
atomic interaction and disorder.
In contrast to other condensed-matter systems
where quantum phase transitions can also take place,
optical lattices provide a unique possibility to {\it control\/} the disorder
which can be created by several methods.
Truly random potentials with continuous disorder distribution can be created using
laser speckles~\cite{Inguscio1,Inguscio2,Aspect,Schulte}
which leads to random contributions to the tunneling amplitudes as well as
on-site energy shifts.
In addition to that, the atomic interaction energies can be made random~\cite{GWSS},
if the lattice loaded by cold atoms is placed near a wire inducing a spatially
random magnetic field~\cite{Schmiedmayer}.
Disorder with discrete probability distribution can be created introducing a second atomic species
strongly localized on random sites~\cite{Vignolo,Castin,HCR}
which leads only to random shifts of the on-site energies.
With the aid of incommensurate lattices
one can make the tunneling amplitudes and the on-site energies
quasi-random~\cite{Lew03,Santos05,FLGFI}.

Until recently, studies of QPTs in cold atoms
were dealing with continuous disorder distributions of different types.
QPTs in the presence of disorder with discrete probability distribution
were studied only for interacting electrons in binary
alloys~\cite{alloys}. The role of this type of disorder in QPTs of cold bosons 
starts to become also a subject of research~\cite{MF}.
In the present work, we shall investigate QPTs of ultracold
bosons in a lattice with disorder which is created by the interaction with impurity
atoms localized at random lattice sites. 
The underlying theoretical model is the Bose-Hubbard Hamiltonian with
random on-site energies according to a binary probability distribution.
The problem was recently addressed in the work by \mbox{Mering} and \mbox{Fleischhauer}~\cite{MF},
who employed the DMRG-method to study the system.
Our analysis 
will be based either on an exact numerical diagonalization
for sufficiently small systems or a Bose-Fermi mapping for the case of
hard-core bosons in a one-dimensional lattice. These exact methods will be
applied to obtain the phase diagram at zero temperature. Moreover,
the role of finite temperature will be studied. In addition, we
evaluate experimentally accessible quantities, such as  the momentum distribution,
the static and dynamic structure factors and the density of excited
states. 

The paper is organized as follows.
In Sec.~\ref{Hamiltonian}, we describe the theoretical model of the
Bose-Hubbard Hamiltonian with binary disorder, studied in the present work.
In Sec.~\ref{SCE}, the physics of the Mott-insulator (MI) phases is explained
and their phase-boundaries are calculated by employing perturbation theory
with respect to the hopping amplitude.
Section~\ref{ED} deals with exact numerical calculations of the many-particle ground
states for small lattices.
In Sec.~\ref{BFM}, we consider a one-dimensional system in the limit of strong interaction,
which is exactly solvable through the Bose-Fermi mapping.
A summary of the work is given in Sec.~\ref{Conclusions}.

\section{\label{Hamiltonian}Hamiltonian}

We consider a system of ultracold interacting bosons in a $d$-dimensional
hypercubic lattice described by the Bose-Hubbard Hamiltonian
\begin{eqnarray}
\label{BHH}
\hat H
&=&
-J
\sum_{
      \langle
      {\bf i},{\bf j}
      \rangle
     }
    \hat a_{\bf i}^\dagger
    \hat a_{\bf j}
+
\frac{U}{2}
\sum_{\bf i}
    \hat a^\dagger_{\bf i}
    \hat a^\dagger_{\bf i}
    \hat a_{\bf i}
    \hat a_{\bf i}
\nonumber\\
&+&
\sum_{\bf i}
\left(
    \epsilon_{\bf i}
    -
    \mu
\right)
\hat a^\dagger_{\bf i}
\hat a_{\bf i}
\;,
\end{eqnarray}
where $J$ is the tunneling matrix element for the nearest lattice
sites, $U$ is the on-site atom-atom interaction energy, and $\mu$ is
the chemical potential.
Throughout the paper, we will be dealing with repulsive interaction, i.e., $U>0$.
We assume periodic boundary conditions. The annihilation and creation operators,
$\hat a_{\bf i}$ and $\hat a^\dagger_{\bf i}$, obey the bosonic commutation relations.

The disorder described by the random terms $\epsilon_{\bf i}$ is assumed to be created
by the presence of impurity atoms located at fixed random positions
(quenched disorder).
This type of disorder is diagonal in the sense that it does not lead
to any contributions to the hopping term.
If it has a ``fermionic" character, i.e.,
if there is at most one impurity at each lattice site,
the probability distribution of on-site energies $\epsilon_{\bf i}$ is given by
\begin{equation}
\label{p}
p(\epsilon)
=
p_0
\delta(\epsilon)
+
(1-p_0)
\delta(\epsilon-U')
\;,\quad
p_0=(L-L')/L
\;,
\end{equation}
with $0\le p_0\le 1$,
where $U'$ is the boson-impurity interaction energy.
$L$ is the number of lattice sites and $L'$ is the number of impurities.
$p_0=1$ or $U'=0$ corresponds to the pure case.
The binary disorder distribution~(\ref{p}) implies that the system under consideration remains
invariant under the transformation $p_0 \to 1-p_0$, $U' \to -U'$.
This can be easily seen if we subtract $U'/2$ from the on-site energies $\epsilon_{\bf i}$,
i.e., if we make the replacement $\epsilon\to\epsilon+U'/2$ on the r.h.s.
of Eq.~(\ref{p}).
Therefore, it is enough to consider the case $U'\ge0$.

\section{\label{SCE}Strong-coupling expansion}

It is known that the Bose-Hubbard model with continuous disorder distribution
possesses a rich phenomenology of phases resulting in a nontrivial
phase diagram \cite{Fisher}. For the present case of binary disorder, we determine
the zero temperature phase diagram next.
The physics of the MI phases can be understood and their phase boundaries
be calculated with a good accuracy by treating
the hopping term in the Hamiltonian~(\ref{BHH}) as a perturbation~\cite{note}.
This can be done in arbitrary spatial dimensions and, in this section,
we will not impose any restrictions with respect to the dimensionality $d$.
For the binary disorder distribution~(\ref{p}) it is convenient
to consider the entire lattice as consisting of two disconnected sublattices
${\cal L}_0$ and ${\cal L}_1$.
The sublattice ${\cal L}_1$ consists of the $L'$ potential wells which are shifted by
$\epsilon_{\bf i}=U'$ with respect to the wells of the other sublattice ${\cal L}_0$.
In what follows, it is assumed that $0<p_0<1$.
One may then define the local chemical potentials of the sublattices
${\cal L}_0$ and ${\cal L}_1$ as $\mu$ and $\mu-U'$, respectively.
The special case of a pure lattice can be retrieved in the limit $U'\to 0$.

\begin{figure}[t]

\hspace{-2cm}  \includegraphics[width=10cm]{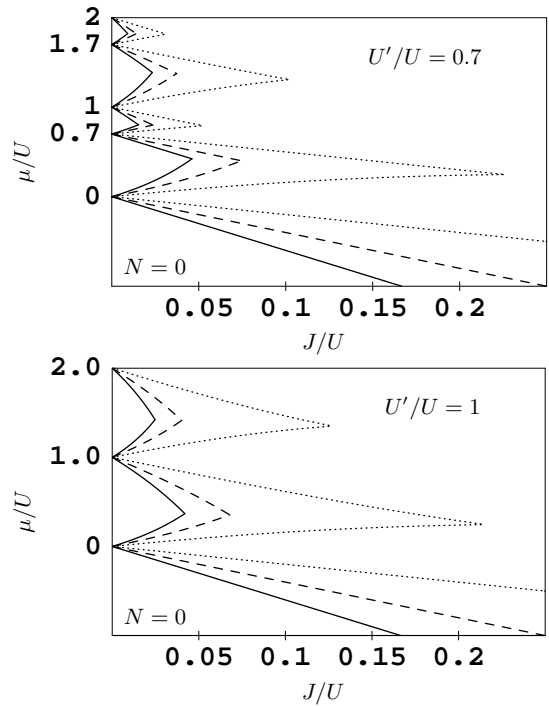}

\caption{
         The boundaries of the MI phases obtained by
         the strong-coupling expansion in $d=1$ (dotted), $2$ (dashed), $3$ (solid)
         for $U/U'=0.7$ (top) and for $U/U'=1$ (bottom).
        }
\label{pdsce}
\end{figure}

In the limit of vanishing hopping ($J=0$),
the states of the system are characterized by the occupation numbers of each lattice site.
In the ground state corresponding to the MI,
we have $n_0$ bosons at each site of the sublattice ${\cal L}_0$,
and $n_1$ bosons at each site of the sublattice ${\cal L}_1$,
where $n_0$ and $n_1$ are the smallest non-negative integers larger than or equal to
$\mu/U$ and $(\mu-U')/U$, respectively.
In general, $n_1\le n_0$ and the total number of bosons $N$ is given by
$N=n_0(L-L')+n_1 L'$.
We are interested in the thermodynamic limit $N\to\infty$, $L\to\infty$, $L'\to\infty$,
where $N/L$ as well as $p_0$ remain finite.
The particle and hole excitations in the limit of infinite lattices are localized
within the Lifshitz rare regions~\cite{Lifshitz}
which consist of infinitely large connected regions
of either sublattice ${\cal L}_0$ or ${\cal L}_1$ depending on the
disorder parameter $U'$ and the chemical potential $\mu$. This allows us to work out
the boundaries of the MI regions by generalizing the method of
strong-coupling expansion developed in Ref.~\cite{FM}.
Due to the infinite extent of the Lifshitz rare regions, the boundaries of the
MI regions should not depend on $p_0$. This dependence appears as a finite size
effect which we do not consider in this section.

The region $\mu<-2dJ$ corresponds to a vanishing particle number $n_0=n_1=0$.
The intervals $n_0-1<\mu/U<n_0$, $n_0=1,\dots,[U'/U]$,
with $[\dots]$ denoting the integer part, correspond to the MI with $n_1=0$.
The lowest-energy particle-hole excitations are
created by transferring one atom among the lattice sites of the sublattice ${\cal L}_0$.
The energy gap for the creation of this excitation equals $U$ at $J=0$.
The MI phases are enclosed in the interval $\mu_h(n_0)<\mu<\mu_p(n_0)$, where
\begin{eqnarray}
\label{muphd}
\mu_p(n)
&=&
U n
-
2dJ(n+1)
\\
&+&
\frac{J^2}{U}
n
\left[
    d(5n+4)
    -
    4d^2(n+1)
\right]
\nonumber\\
&+&
\frac{J^3}{U^2}
n(n+1)
\left[
    -8d^3(2n+1)
    +
    d^2(25n+14)
\right.
\nonumber\\
&&
\hspace{3cm}
\left.
    -
    4d(2n+1)
\right]
+
O
\left(
    J^4
\right)
\;,
\nonumber\\
\mu_h(n)
&=&
U(n-1)
+
2dJn
\nonumber\\
&-&
\frac{J^2}{U}
(n+1)
\left[
    d(5n+1)
    -
    4d^2n
\right]
\nonumber\\
&+&
\frac{J^3}{U^2}
n(n+1)
\left[
    8d^3(2n+1)
    -
    d^2(25n+11)
\right.
\nonumber\\
&&
\hspace{3cm}
\left.
    +
    4d(2n+1)
\right]
+
O
\left(
    J^4
\right)
\;,
\nonumber
\end{eqnarray}
are the boundaries of the MI phase in the pure case~\cite{FM}.

The next intervals $n_0-1<\mu/U<n_0$, where $n_0=[U'/U]+1,\dots$,
are split into two subintervals.
In the lower subinterval $n_0-1<\mu/U<n_0-1+\left\{U'/U\right\}$,
where $\left\{\dots\right\}$ denotes the fractional part, $n_1=n_0-[U'/U]-1$.
The lowest-energy particle-hole excitation of this state can be created by
transferring one atom from the sublattice ${\cal L}_0$ to the sublattice ${\cal L}_1$.
As a result of this transfer, the energy of the initial state is increased by
$U\left\{U'/U\right\}$.
Perturbative calculations in the thermodynamic limit show that
it is located in the interval $\mu_h({n_0})<\mu<U'+\mu_p({n_1})$ of the phase diagram (Fig.~\ref{pdsce}).

In the upper subinterval $n_0-1+\left\{U'/U\right\}<\mu/U<n_0$, $n_1=n_0-[U'/U]$.
The lowest-energy particle-hole excitation of this state can be created by
transferring one atom from the sublattice ${\cal L}_1$ to the sublattice ${\cal L}_0$.
As a result of this transfer, the energy of the initial state is increased by
$U-U\left\{U'/U\right\}$.
According to the perturbation theory,
it is located in the interval $U'+\mu_h({n_1})<\mu<\mu_p({n_0})$,
where $\mu_h(n)$ and $\mu_p(n)$ are given by Eq.~(\ref{muphd}).

The two subintervals exist only if $\left\{U'/U\right\}$ does not vanish,
otherwise the lower subinterval disappears. The upper one then becomes extended from
$\mu=(n_0-1)U$ to $\mu=n_0 U$. In this case, it is energetically more favorable
to create the particle-hole excitations by transferring one atom among the lattice sites
of the sublattice ${\cal L}_0$ and the MI phase is enclosed within the interval
$\mu_h({n_0})<\mu<\mu_p({n_0})$.

MI phases with equal occupation numbers $n_0=n_1=n$ exist only for
$0\le U'/U < 1$, i.e., $[U'/U]=0$ and $\left\{U'/U\right\}=U'/U$.
They are located within the intervals $n_0-1+U'/U<\mu/U<n_0$.

The MI regions for $U'/U=0.7$ and for $U'/U=1$ are shown in Fig.~\ref{pdsce}.
In the case $U'/U=0.7$, we have only split MI regions with $n_0=1,2,\dots$,
where the lower and upper parts correspond to $n_1=n_0-1$ and $n_1=n_0$, respectively.
In the case $U'/U=1$, there are only nonsplit MI regions with $n_1=n_0-1$.
With the increase of the dimensionality $d$, the MI regions become smaller.

Since we are dealing with a disordered system, one can expect the existence
of the Bose-glass (BG) phase~\cite{Fisher}. However, the method of strong-coupling expansion
in its present form does not allow us to detect the corresponding regions on
the phase diagram. It does not give the opportunity to investigate
the temperature-dependent effects either.
Therefore, other methods are needed in order to study the complete
phase diagram of the system.
They are employed in the next sections, where we consider only one-dimensional
lattices.

\section{\label{ED}Exact diagonalization}

In this section we study zero-temperature properties of the system by means of
exact numerical diagonalization of
the Bose-Hubbard Hamiltonian along the lines of Ref.~\cite{RB}.
For this we determine first the boundaries of the regions
in the $(\mu,J)$ plane corresponding to different total particle numbers $N$.
This requires
calculations of the ground-state energies $E_N$ of the Hamiltonian~(\ref{BHH})
for different $N$ and can be done exactly with the aid
of iterative numerical solvers for sparse matrices of large dimensions.

\begin{figure}[t]
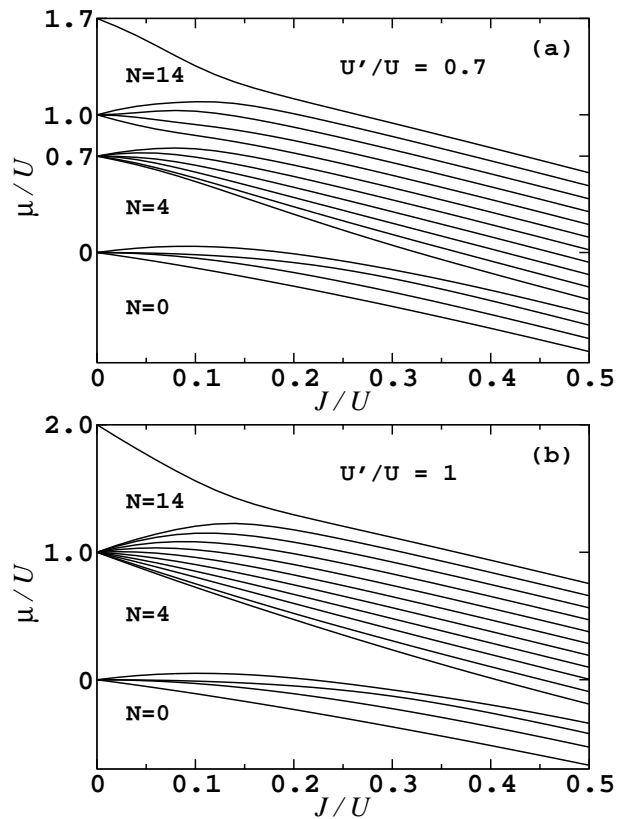


  \includegraphics[width=8cm]{fig2a.eps}

  \includegraphics[width=8cm]{fig2b.eps}

\caption{
         The regions with occupation numbers $N=0,\dots,14$ obtained by
         the exact diagonalization
         of the Hamiltonian~(\ref{BHH}) for $L=10$, $L'=6$ for one
	 disorder realization characterized by the Fock state 
	 $|0011101011\rangle$ for $U'/U=0.7$ (a) and for $U'/U=1$
	 (b).
        }
\label{pde}
\end{figure}

The results of these calculations for a small one-dimensional lattice
are presented in Fig.~\ref{pde}.
The solid lines indicate the boundaries $\mu_N=E_N-E_{N-1}$ of the regions with different
occupation numbers $N$ of the lattice.
These calculations are performed for $L=10$ and $L'=6$ and for one disorder realization,
where the spatial distribution of impurities is described by the Fock state
$
|0011101011\rangle
$.
The maximal number of bosons is $15$.
The lowest line is the boundary between $N=0$ and $N=1$,
the next one is the boundary between $N=1$ and $N=2$ and so on.

In the case $U'/U=0.7$~(Fig.~\ref{pde}a), the regions with the occupation numbers
$N=L-L'=4$, $N=L=10$, $N=2L-L'=14$ appear to be larger than
the others indicating that there are MI phases for these occupations.
In spite of the large contribution of the finite-size effects,
the shape of these regions is in good agreement with the results
of the strong-coupling expansion, see Fig.~\ref{pdsce}.

As it was discussed in the previous section,
the MI phases with integer filling factors disappear if $U' \ge U$
and only the MI phases with incommensurate fillings remain.
This behavior can be also seen in Fig.~\ref{pde}b.

With the increase of the number of lattice sites $L$,
the boundaries of the regions with different occupation number are 
closer to each other and in the thermodynamic limit they should densely cover
the whole $(\mu,J)$ plane except the MI regions. In order to really see this
transition to the thermodynamic limit as well as to determine the boundaries
of the MI regions, one has to  vary
the number  $L$ of lattice sites and the number $N$ of bosons 
in a wide range which is difficult
here because the dimension of the bosonic Hilbert space grows exponentially
with $N$ and $L$.

The fraction $f_s^N$ of the total atom number $N$ which are in the
superfluid phase (superfluid fraction) 
can be calculated with the aid of Peierls phase factors. They have to
be introduced in the Hamiltonian~(\ref{BHH}) by means of the replacement
$\hat a_{i}^\dagger\hat a_{i+1}\to\hat a_{i}^\dagger\hat
a_{i+1}e^{i\phi}$. By calculating the free energy $F_N(\phi)$ for some 
small value of $\phi$, the superfluid fraction is determined as~\cite{defsf,RB}
\begin{equation}
\label{fs}
f_s^N
=
\lim_{\phi\to0}
\frac{F_N(\phi)-F_N(0)}{J \phi^2 N}
\;.
\end{equation}
The limit $\phi\to0$ can be calculated exactly if the complete solution of
the eigenvalue problem is known (see Ref.~\cite{RB} and the discussion in Sec.~\ref{BECSF}).
However, all the numerical solvers for large sparse matrices allow efficient
calculations only of a small number of the eigenstates. This is the reason why
the superfluid fraction defined by Eq.~(\ref{fs}) is usually worked out for some
small but nonvanishing value of $\phi$.

\begin{figure}[t]

\hspace{-2cm}  \includegraphics[width=10cm]{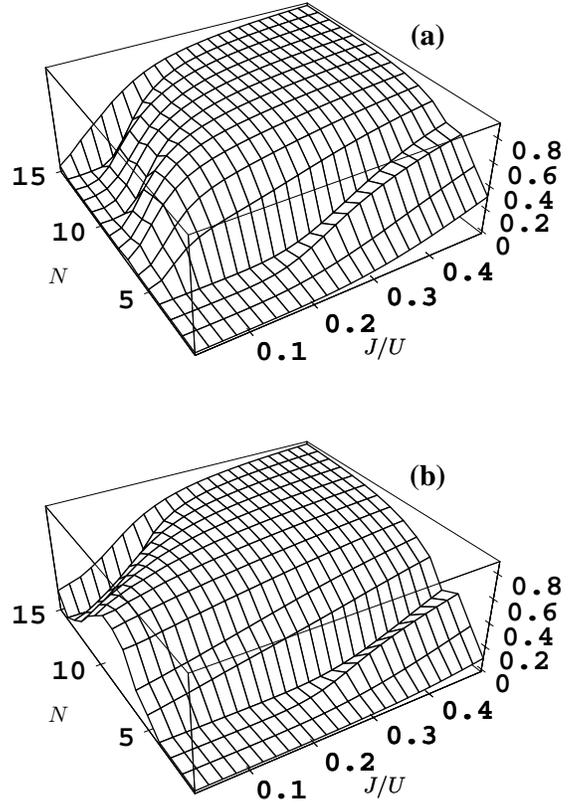}

\caption{
         Superfluid fraction $f_s^N$ vs $N$ and $J$ obtained by exact diagonalization
         for $L=10$, $L'=6$ for $U'/U=0.7$ (a) and for  $U'/U=1$ (b).
         The points for different values of $N$ are connected in order
         to guide eyes.
        }
\label{sf}
\end{figure}

The behavior of $f_s^N$ for different $N$ is shown in Fig.~\ref{sf}
for the same fixed disorder realization as above.
In these calculations, $\phi=0.01$.
In the case $U'/U=0.7$, the superfluid fraction vanishes not only for
fillings which allow the existence of the MI phases ($N=4,10,14$)
but also for all the others if the hopping parameter $J$ is small enough.
This suggests a phase transition from the superfluid into the BG phase.
The case $U'/U=1$ looks different. The superfluid fraction vanishes only
for $N=1,2,3,4,14$ and remains finite for all the others, even for small values
of $J/U$. This suggests that the BG phase exists
in the extended regions of the $(\mu,J)$ plane corresponding to low fillings
but is strongly suppressed for higher fillings.

Since exact diagonalization can be performed only for very small lattices,
it is difficult to provide a satisfactory description of the phase diagram of the system.
Finite-size effects can be much better controlled
in the case of hard-core bosons, which is treated in the next section.

\section{\label{BFM}Bose-Fermi mapping}

We consider the hard-core limit of infinitely strong repulsion 
$U\to\infty$ in a one-dimensional lattice,
which is exactly solvable via the Jordan-Wigner transformation~\cite{JW,LSM}
\begin{equation}
\label{JWT}
\hat a_l
=
\exp
\left(
    i \pi
    \sum_{j<l}
    \hat c_j^\dagger \hat c_j
\right)
\hat c_l
\end{equation}
where $\hat c_l$ and $\hat c^\dagger_l$
are the fermionic annihilation and creation operators.
Under this transformation the Hamiltonian~(\ref{BHH}) takes the form
\begin{equation}
\label{HHC}
\hat H
=
-J
\sum_{i=1}^L
\left(
    \hat c_i^\dagger
    \hat c_{i+1}
    +
    \hat c_{i+1}^\dagger
    \hat c_i
\right)
+
\sum_{i=1}^L
\left(
    \epsilon_i
    -
    \mu
\right)
\hat c^\dagger_i
\hat c_i
\;.
\end{equation}
Periodic boundary conditions for bosons are equivalent to the requirement
\begin{equation}
\hat c_{L+1}
=
\exp
\left(
    - i \pi
    \sum_{j=1}^L
    \hat c_j^\dagger \hat c_j
\right)
\hat c_1
\;,
\end{equation}
which implies periodic boundary conditions for fermions if the number of particles
$N$ is odd, otherwise one should use the corresponding 
antiperiodic boundary conditions in the
Hamiltonian~(\ref{HHC}).

The $N$-particle eigenstates of the Hamiltonian~(\ref{HHC}) can be constructed
from the $L$ single-particle eigenstates as 
\begin{equation}
|\alpha\rangle
=
\sum_{i=1}^L
\varphi_\alpha(i)
{\cal T}^{i-1}
|\underbrace{1 0 \dots 0}_L\rangle
\end{equation}
with the eigenenergies $\varepsilon_\alpha$, $\alpha=1,\dots,L$.
Here, ${\cal T}$ is the translation operator.
Its action ${\cal T}^i$ on the one-particle Fock state
$|1 0 \dots 0\rangle$
results in the shift of the particle's position by $i$ lattice sites.
This allows to treat much larger lattices compared to the case of soft-core bosons
considered in Sec.~\ref{ED}. However, one still cannot avoid numerics
because the analytical solution of the single-particle eigenvalue problem in the presence
of disorder is not known.

In the infinite-$U$ limit, the occupation  numbers of the individual lattice sites can 
be $0$ or $1$ implying that the maximal number of atoms $N$ cannot be larger than $L$.
The state with $N=L$ is always a MI as no hopping can take place any
more and a non-trivial treatment of the MI phase 
in the hard-core limit is possible only for $N<L$.

\subsection{\label{BECSF}Bose-Einstein condensation and superfluidity}

Before starting the discussion of the remaining  part of the phase diagram, 
some remarks on the Bose-Einstein condensation (BEC) and superfluidity of hard-core bosons
in 1D are in order. In the absence of disorder, the spatial correlation function
$\langle \hat a_i^\dagger \hat a_j\rangle$ in the limit $|i-j|\to\infty$
decays as $|i-j|^{-1/2}$. The presence of disorder makes this decay faster, i.e.,
there is no off-diagonal long-range order and BEC \cite{MTEG}.

The absence of BEC does not exclude in general the superfluidity.
As it was shown in Ref.~\cite{GS} for the Gaussian disorder,
the system of one-dimensional soft-core bosons is in the delocalized (superfluid)
state, if the correlation function $\langle\hat a_i^\dagger \hat a_j\rangle$
for large distances decays slower than $|i-j|^{-1/3}$ which leads to the divergence
of the localization length. Since the correlation function of hard-core bosons with disorder
decays faster than $|i-j|^{-1/2}$, the criterion of Ref.~\cite{GS} excludes the existence
of the superfluid phase. The question is whether this remains true for the
binary disorder.

The superfluid fraction is defined by Eq.~(\ref{fs}). The limit $\phi\to0$ can be calculated
 making use of the perturbative approach of Ref.~\cite{RB}.
For hard-core bosons in 1D, this leads to the following general expression at $T=0$:
\begin{eqnarray}
\label{fshc}
f_s^N
&=&
\frac{1}{2N}
\sum_{i=1}^L
\sum_{\alpha=1}^N
\left[
    \varphi_\alpha^*(i)
    \varphi_\alpha(i+1)
    +
    {\rm c.c.}
\right]
\\
&-&
\frac{J}{N}
\sum_{\alpha=N+1}^L
\sum_{\beta=1}^N
\frac
{1}
{\varepsilon_\alpha-\varepsilon_\beta}
\nonumber\\
&\times&
 \left|
     \sum_{i=1}^L
     \left[
         \varphi_\alpha^*(i)
         \varphi_\beta(i+1)
         -
         \varphi_\alpha^*(i+1)
         \varphi_\beta(i)
     \right]
 \right|^2
\;,
\nonumber
\end{eqnarray}
where $\varphi_\alpha(L+1)=(-1)^{N+1}\varphi_\alpha(1)$.
In the clean case, the second term in Eq.~(\ref{fshc}) vanishes and
in the thermodynamic limit we get $f_s=\frac{\sin\pi n}{\pi n}$, where $n=N/L$~\cite{com}.
This result does not depend on $J$.

\begin{figure}[t]
\centering


  \includegraphics[width=8cm]{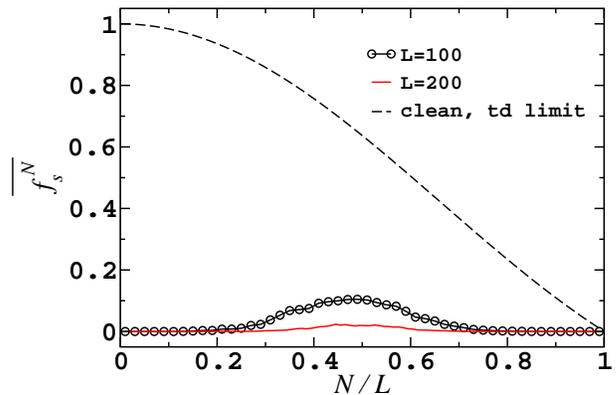}

\caption{
         Superfluid fraction for $J/U'=1$, $L'=L/2$, $L=100$ (circles), $200$ (solid line)
         averaged over $400$ disorder realizations.
         The dashed line shows the superfluid fraction in the thermodynamic limit without
         disorder, i.e., $f_s=\frac{\sin(\pi N/L)}{\pi N/L}$, see text.
        }
\label{sfh}
\end{figure}

The results of numerical calculations for the binary disorder are shown in Fig.~\ref{sfh}
for $J/U'=1$ for two different lattice sizes.
Here and later on the overline indicates the disorder-averaged quantity.
For a small lattice size $L=100$, $f_s$ appears to be finite.
However, when increasing $L$ to $L=200$, $f_s$ approaches zero.
For smaller values of $J/U'$, smaller
lattice sizes $L$ are sufficient in order to see that $f_s$ indeed vanishes
in the thermodynamic limit, i.e., we expect only insulating phases in
our system. For comparison, the corresponding result for the clean
case is also shown in Fig.~\ref{sfh}, see dashed line. 

Since the hard-core limit works for $U'/U<1$ and $J/U\ll 1$,
this qualitatively agrees with the behavior of $f_s^N$ for $N\le L=10$ and relatively small
$J/U$ calculated by means
of exact diagonalization in the case of soft-core bosons
(Fig.~\ref{sf}a).

\subsection{Quantum phases}

\begin{figure}[t]

\hspace{-2cm}  \includegraphics[width=10cm]{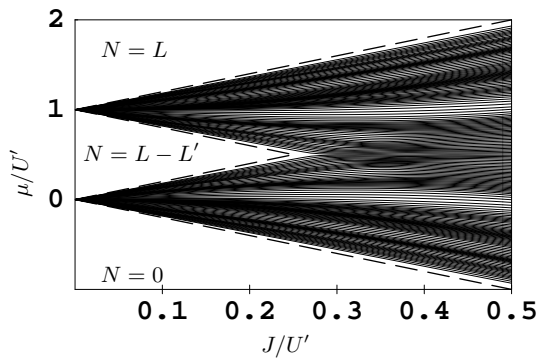}

\caption{
         The regions with occupation numbers $N=0,\dots,L$ obtained by
         the exact diagonalization of the Hamiltonian~(\ref{HHC}) for $L=200$, $L'=100$.
         The results are averaged over $400$ disorder realizations.
         Dashed lines are analytical results for the boundaries of the MI regions
         in the thermodynamic limit.
        }
\label{pdhc}
\end{figure}

The boundaries of the MI-regions at zero temperature can be easily calculated analytically
in the thermodynamic limit following the general treatment of Sec.~\ref{SCE}
for $d=1$, $U'<U$. The region $\mu<-2J$ contains no atoms.
The MI phases with non-integer filling factors are located in the interval
$\mu_h({n_0})<\mu<U'+\mu_p({n_1})$, where $n_0=1$, $n_1=0$. In the limit $U\to\infty$,
all the terms in Eq.~(\ref{muphd}) proportional to $J^p$ with $p>1$ vanish and we get explicitly $2J<\mu<U'-2J$, where $J<U'/4$.
$U'+\mu_h({n_1})<\mu<\mu_p({n_0})$ with $n_0=n_1=1$, i.e., $U'+2J<\mu<\infty$
corresponds to the MI with $N=L$.
These results are shown in Fig.~\ref{pdhc} by dashed lines.

The same boundaries as well as the boundaries of the regions with different occupation
numbers $\mu_N$ can be calculated numerically by means of exact diagonalization.
$\mu_N$ are equal to
the single-particle eigenenergies $\varepsilon_N$ because the ground-state energy of $N$
non-interacting fermions is the sum of $N$ lowest single-particle eigenenergies.
The region $\mu<\varepsilon_1$ contains no atoms.
The MI phase with $L-L'$ atoms is located in the interval
$\varepsilon_{L-L'}<\mu<\varepsilon_{L-L'+1}$.
If $\mu>\varepsilon_L$, we have the MI with $N=L$.
The boundaries $\varepsilon_1,\dots,\varepsilon_L$
are shown by solid lines in Fig.~\ref{pdhc} for $L=200$.
The lines $\varepsilon_1$, $\varepsilon_{L-L'}$, $\varepsilon_{L-L'+1}$, $\varepsilon_L$
are outside of the corresponding regions determined in the thermodynamic limit,
similar to what is seen in Fig.~\ref{pde}.
They come closer to the results obtained in the limit $L\to\infty$
if the lattice size $L$ is increased.

\begin{figure}[t]
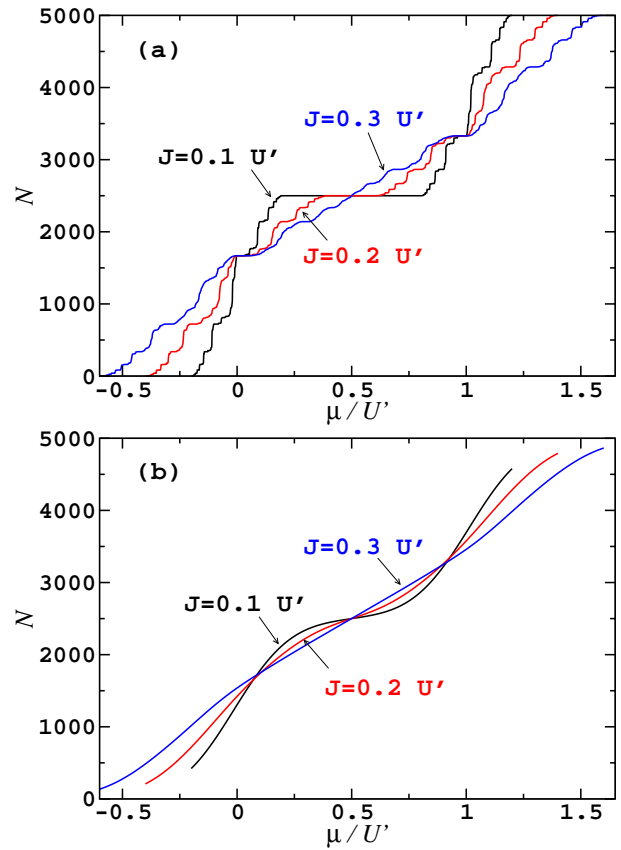

\centering


  \includegraphics[width=8cm]{fig6a.eps}

  \includegraphics[width=8cm]{fig6b.eps}

\caption{(Color online)
         $N(\mu)$ in the hard-core limit for $k_BT/U'=0$ (a), $0.1$ (b)
         and for one disorder realization which is the same in (a) and (b).
         $L=5000$, $L'=2500$,
         $J/U'=0.1$ (black), $0.2$ (red), $0.3$ (blue).
        }
\label{nmu}
\end{figure}

The distribution of lines $\mu_N=\varepsilon_N(J)$, $N=1,\dots,L$, in Fig.~\ref{pdhc}
is very inhomogeneous which leads
to the fact that the compressibility of the system varies in a rather wide range.
This characteristic feature remains preserved for larger lattices as well and it is easier
to see it in the behavior of $N(\mu)$ which is given by
\begin{equation}
N(\mu)
=
\sum_{\alpha=1}^L
f(\varepsilon_\alpha)
\;,\quad
f(\varepsilon)
=
\frac{1}
{
 \exp
 \left[
     (\varepsilon-\mu)/k_B T
 \right]
 +
 1
}
\;.
\label{nofmudef}
\end{equation}
The plots $N(\mu)$ at $T=0$ are shown in Fig.~\ref{nmu}a.
The central plateaus ($N=L-L'$) around $\mu/U'=0.5$ which exist for $J/U'<0.25$
correspond to the MI phase. Quasi-plateaus of $N(\mu)$ which exist at any values
of $J/U'$ correspond to the regions on the $(J,\mu)$-diagram with low density of lines
$\mu_N(J)$.

\begin{figure}[t]
\centering





  \includegraphics[width=8cm]{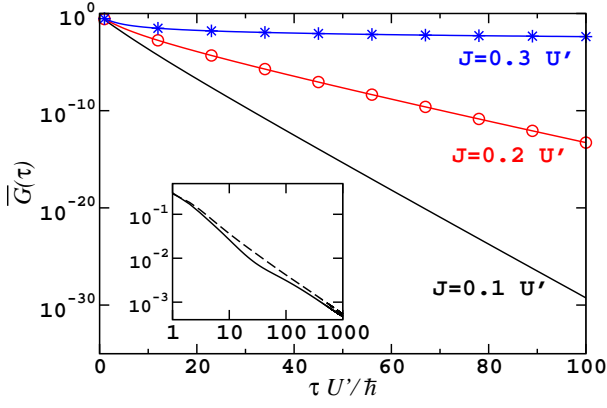}

\caption{(Color online) 
         Linear-log plot of $G(\tau)$ for $L=400$ averaged over $5000$
	 disorder realizations for
         $N=200$ and $J/U'=0.1$ (black solid), $J/U'=0.2$ (red circles) and $J/U'=0.3$ 
	 (blue stars).
         Inset: same in a log-log plot for $J/U'=0.3$ for  $N=132$
	 (solid) and  $N=200$ (dashed)
         which corresponds to $\mu/U'\approx 0$ and $0.5$, respectively.
        }
\label{G}
\end{figure}

In order to clarify the physical interpretation of different parts of the phase diagram,
we have calculated the time-dependent Green's function
$G(\tau)=\langle \hat a_i(\tau) \hat a_i^\dagger(0)\rangle$ defined for $\tau>0$
which determines the density of states of the single-particle excitations as well as
the superfluid susceptibility~\cite{Fisher}.
The plots of $G(\tau)$ as a function of  the imaginary time $\tau$ are shown in Fig.~\ref{G}.
In the MI phase, $G(\tau)$ at large $\tau$ is an exponential function
of $\tau$. This also holds  for $N=L-L'$, $J/U'<0.25$.
If the number of particles remains the same but $J/U'$ is increased, the exponential
decay of $G(\tau)$ is replaced by the $1/\tau$-law indicating that now we are in
the BG phase (Fig.~\ref{G} main) according to Ref.~\cite{Fisher}.
$1/\tau$-law is also observed for other particle numbers corresponding to the
quasi-plateaus of $N(\mu)$ (inset of Fig.~\ref{G}) which allows to
interpret them as belonging to the BG phase, 
in spite of the fact that the compressibility is extremely small.

The density of states for the single-particle excitations can be determined in terms of
the Fourier transformed single-particle Green's function $\tilde G(E)$ as
$\rho(E) = - \frac{1}{\pi}\, {\rm Im}\, \tilde G(E)$~\cite{AGD}.
For hard-core bosons, it takes the form
\begin{equation}
\label{doshc}
\rho(E)
=
\sum_{\alpha,\beta}
f(\varepsilon_\alpha)
\left[
    1 - f(\varepsilon_\beta)
\right]
\delta
\left(
    E-\varepsilon_\beta+\varepsilon_\alpha
\right)
\;,
\end{equation}
where $\varepsilon_\beta-\varepsilon_\alpha$ are the energies of single-particle excitations
caused by the transfer of one particle from the energy level
$\varepsilon_\alpha$ to the energy level $\varepsilon_\beta$.

\begin{figure}[t]
\centering

\psfrag{e}[c]{$E/U'$}
\psfrag{dos}[b]{$\overline{\rho}(E)$}

  \includegraphics[width=8cm]{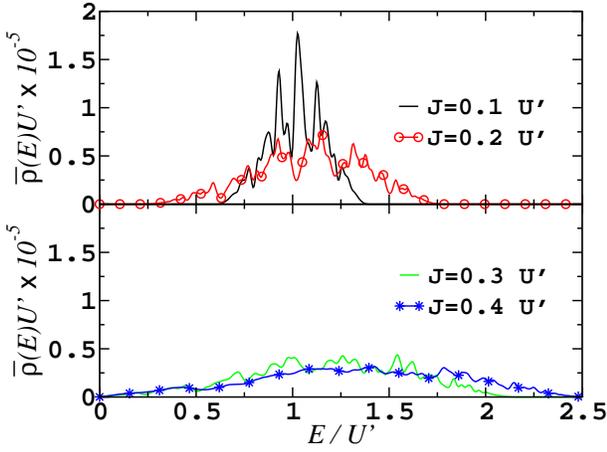}

\caption{(Color online)
         Density of excited states at $T=0$ averaged over $200$ disorder realizations.
         $L=400$, $L'=200$, $N=200$,
         $J/U'=0.1$ (black), $0.2$ (red), $0.3$ (green), $0.4$ (blue).
        }
\label{dos}
\end{figure}

Numerical calculations were performed assuming that the energy levels
have a finite lifetime.
The $\delta$-function is approximated by a Gaussian with the standard
deviation $0.01U'$. The disorder-averaged energy dependences at $T=0$ are shown in Fig.~\ref{dos} for different values of $J$.
$\overline{\rho}(E)$ has always a multi-maxima structure which stems from the inhomogeneous distribution
of the single-particle eigenenergies (see Figs.~\ref{pdhc},~\ref{nmu}a).

\begin{figure}[t]
\centering


\hspace{-2cm}  \includegraphics[width=10cm]{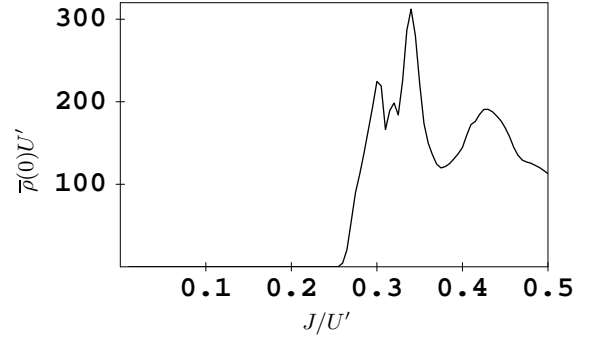}

\caption{
         Density of excited states at $T=0$ and $E=0$ averaged over $200$ disorder
         realizations.
         $L=400$, $L'=200$, $N=200$.
        }
\label{dos0}
\end{figure}

The density of excited states at $E=0$ as a function of $J$ is shown in Fig.~\ref{dos0} for $N=L-L'$.
$\rho(0)$ vanishes for $J/U'\lesssim 0.25$, and is different from zero otherwise.
For other particle numbers, $\rho(0)$ does not vanish for any $J$.
This is directly related to the asymptotic properties of the time-dependent Green's function
$G(\tau)$ discussed above.
$\rho(0)>0$ is a clear-cut signature of the BG~\cite{Fisher,KPG}.

At finite temperature, $N(\mu)$ defined in Eq.~(\ref{nofmudef}) 
becomes a smooth function (Fig.~\ref{nmu}b).
The MI plateaus which are clearly seen for $J/U'=0.1,0.2$ at $T=0$ (Fig.~\ref{nmu}a)
are smeared out and the quasi-plateaus disappear even at rather small $T$ indicating that
the compressibility never vanishes.
Nevertheless, one can say that the MI still exists as long as the gap for the creation
of particle-hole excitations, which is $U'-4J$ in the case of hard-core bosons in 1D,
is larger than the thermal energy $k_B T$. This suggests that the upper and lower boundaries
of the MI-region are given by $U'-2J-k_B T/2$ and $2J+k_B T/2$, respectively,
and the crossover line for the MI on the $(J,T)$-diagram is given by $k_B T_c^{MI}=U'-4J$.

If   temperature is increased, quantum statistics will become less important
and one can expect that the MI phase as well as the BG phase is
destroyed. Then,  a crossover into the normal-gas state is
expected. Since the thermodynamic
properties of hard-core bosons are equivalent to those of the ideal Fermi gas,
the crossover temperature coincides with the Fermi temperature modified
by the disorder.

\subsection{Experimentally measurable quantities}

The Bose-Fermi mapping allows very detailed investigations of different physical
properties of the system which can be directly measured in experiments.
We consider first the momentum distribution~\cite{KPS}
\begin{displaymath}
\langle
\tilde \psi^\dagger (k)
\tilde \psi (k)
\rangle
=
\left|
    \tilde W(k)
\right|^2
\frac{1}{N}
\sum_{l,l'}
\exp
\left[
    i k a (l-l')
\right]
\langle
    \hat a_l^\dagger \hat a_{l'}
\rangle
\;,
\end{displaymath}
where $k$ is the wavenumber, $a$ is the lattice constant, and 
$\tilde W(k)$ is the Fourier transform of the Wannier function $W(x)$ for the lowest
Bloch band of the lattice potential.
The matrix elements
$\langle\hat a_l^\dagger \hat a_{l'}\rangle$
of the $L \times L$ one-particle density matrix
can be worked out for $l>l'$ as a determinant of the
$(l-l')\times(l-l')$ Toeplitz matrix
$G^{(l,l')}$~\cite{LSM,MTEG} as 
\begin{equation}
\langle\hat a_l^\dagger \hat a_{l'}\rangle
=
2^{l-l'-1}
\det G^{(l,l')}
\;.
\end{equation}
The matrix elements of $G^{(l,l')}$ are given by
\begin{equation}
G^{(l,l')}_{i,j}
=
\langle
    c_{l-j+1}^\dagger
    c_{l-i}
\rangle
-
\frac{1}{2}
\delta_{j,i+1}
\;.
\end{equation}
The expectation values
$\langle c_i^\dagger c_j \rangle$
can be calculated using the solution of the single-particle eigenvalue problem as
\begin{equation}
\langle c_i^\dagger c_j \rangle
=
\sum_{\alpha=1}^L
\varphi^*_\alpha(i)
\varphi_\alpha(j)
f(\varepsilon_\alpha)
\;.
\end{equation}

\begin{figure}[t]
\centering



  \includegraphics[width=8cm]{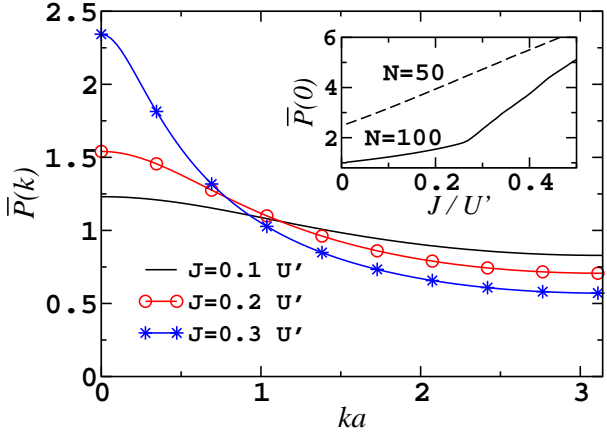}

\caption{(Color online) 
         Momentum distribution in the hard-core limit
         at $T=0$ averaged over $200$ disorder realizations for 
         $L=200$, $L'=100$, $N=100$.
         Inset: Maximum of the momentum distribution.
        }
\label{md}
\end{figure}

The momentum distributions obtained by numerical calculations for $N=L-L'$
are shown in Fig.~\ref{md}. In general,
$
P(k)
=
\langle\tilde\psi^\dagger(k)\tilde\psi(k)\rangle
/|\tilde W(k)|^2
$
is an even and periodic function of $ka$ with the period $2\pi$.
It takes maximal (minimal) values at $ka=\pi m$, $m=0,\pm 2,\pm 4,\dots$
($m=\pm 1,\pm 3,\dots$).
With the decrease of the hopping parameter $J$ the spatial correlations of bosons
become weaker which leads to the broadening of the momentum distribution.
As discussed above, for $N=L-L'$ the system undergoes a phase transition
from the BG to MI, where the spatial correlation functions obey
power and exponential laws, respectively. Therefore, the dependence of the
momentum distribution on $J$ is expected to be weaker in the MI than in the BG,
which is demonstrated in Fig.~\ref{md}.
The transition point is seen as a kink in the
$J$-dependence of the momentum distribution function at $k=0$ (inset of Fig.~\ref{md}).
For $N=(L-L')/2$, which 
always corresponds to the BG phase,
the dependence is almost linear without any kink.

\begin{figure}[t]
\centering

  \includegraphics[width=8cm]{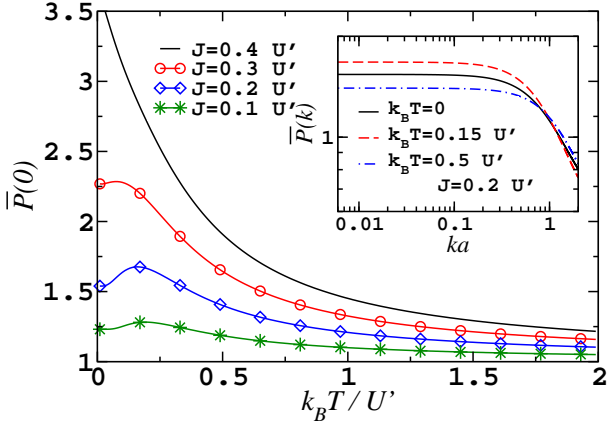}

\caption{(Color online) 
         Temperature dependence of $\bar P(0)$ for 
	 $L=200$, $L'=100$, $N=100$ averaged over 
	 $250$ disorder realizations.
         Inset: Log-log plot of $\bar P(k)$ for the same parameters.
        }
\label{mdt}
\end{figure}

At finite temperature, the qualitative form of the momentum distribution
remains unchanged. As  shown in Fig.~\ref{mdt}, $\bar P(0)$ is a decreasing function
of $T$ for large enough $J/U'$ corresponding to the BG region
in the phase diagram at $T=0$. The largest choice $J=0.4 U'$
is in the BG-region rather far from  the  BG-MI transition point
implying that the averaged momentum distribution has a tendency to become broader due to
the influence of the thermal fluctuations.
For smaller values of $J/U'$, i.e., closer to the 
BG-MI transition, $\bar P(0)$ first increases for small $T$ and then 
decreases further. For the smallest value $J=0.1 U'$, the system is deep
in the MI phase at $T=0$, and the intermediate maximum in
$\bar P(0)$ as a function of $T$ vanishes again.

Useful information about the state of the many-body system can be obtained with
the aid of Bragg spectroscopy~\cite{PS,Esslinger}.
The response of the system to this kind of measurement
is described by the dynamical structure factor, which is defined as
\begin{equation}
S(k,\omega)
=
\int_{-\infty}^\infty
dt
\langle
    \Delta\tilde \rho(k,0)
    \Delta\tilde \rho(-k,t)
\rangle
\exp(-i \omega t)
\;,
\end{equation}
where $\Delta\tilde \rho(k)$ is the spatial Fourier transform of the density-fluctuation operator.
In the case of deep lattices it takes the form
\begin{equation}
\Delta\tilde \rho(k)
=
I_0(k)
\sum_{l}
\left(
    a_l^\dagger a_l
    -
    \langle
         a_l^\dagger a_l
    \rangle
\right)
\exp(ikal)
\;,
\end{equation}
where
\begin{equation}
I_0(k)
=
\int_{-\infty}^\infty
dx
\exp(ikx)
\left|
    W(x)
\right|^2
\;
\end{equation}
with $W(x)$ being the Wannier function for the lowest Bloch band.
The dynamical structure factor for hard-core bosons can be expressed in terms of the
single-particle eigenmodes as~\cite{VM}
\begin{eqnarray}
\label{dsfhc}
S(k,E)
&=&
\hbar
\left|
    I_0(k)
\right|^2
\sum_{\alpha,\beta}
\left|
    \sum_l
    \varphi_\alpha^*(l)
    \varphi_\beta(l)
    e^{ikal}
\right|^2
\nonumber\\
&\times&
f(\varepsilon_\alpha)
\left[
    1 - f(\varepsilon_\beta)
\right]
\delta
\left(
    E-\varepsilon_\beta+\varepsilon_\alpha
\right)
\;.
\end{eqnarray}
This formula resembles Eq.~(\ref{doshc}) for the density of excited states
but has a more complicated structure due to the explicit dependence on the
mode-functions $\varphi_\alpha(l)$.

\begin{figure}[t]
\centering

\psfrag{e}[c]{$E/U'$}
\psfrag{dsf}[b]{$\overline{S(k,E)}/\left|I_0(k)\right|^2$}

  \includegraphics[width=8cm]{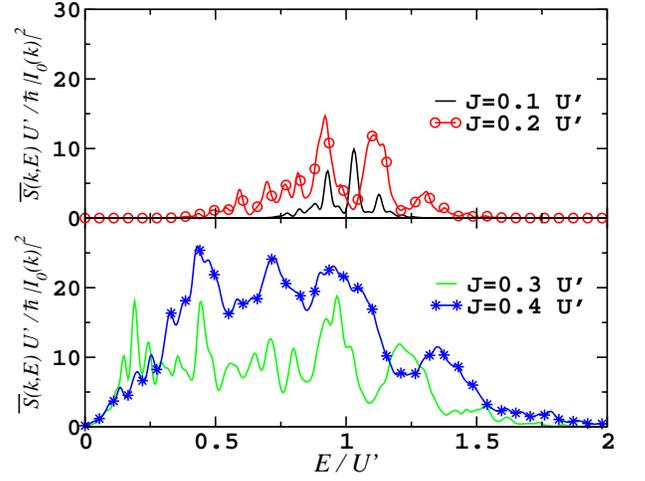}

\caption{(Color online)
         Dynamical structure factor in the hard-core limit at $T=0$
         averaged over $600$ disorder realizations. Parameters are 
         $L=200$, $L'=100$, $N=100$, $ka=\pi/3$.
         Moreover,  $J/U'=0.1$ (black), $0.2$ (red), $0.3$ (green), $0.4$ (blue).
        }
\label{dsf}
\end{figure}

The resulting dependence of the disorder-averaged dynamical structure
factor $\bar S(k,E)$ on energy is shown in Fig.~\ref{dsf} for
$ka=\pi/3$. 
The behavior of $\bar S(k,E)$ is completely different compared to the case of homogeneous
lattices studied in Ref.~\cite{PRB}.
It vanishes in the finite interval of $E$ near zero, provided that $J/U'<0.25$, due to the
energy gap in the excitation spectrum of MI.
With the increase of $J/U'$ the gap decreases. It disappears completely if $J/U'>0.25$
due to the transition into the BG phase.
$\bar S(k,E)$ is broader in the BG phase ($J/U'=0.3,0.4$) than in the MI phase ($J/U'=0.1,0.2$).
Its multi-peak structure is qualitatively related to the density of excited states,
which is shown in Fig.~\ref{dos}.
However, the detailed form of the energy dependence of $S(k,E)$,
which remains preserved for larger lattices as well, is different from $\rho(E)$
due to the nontrivial contributions of the eigenfunctions $\varphi_\alpha(i)$
in Eq.~(\ref{dsfhc}).
The same features are observed for other values of $ka$.

\begin{figure}[b]
\centering

\psfrag{J}[c]{$J/U'$}
\psfrag{ssf}[b]{$\overline{S_0(k)}/\left|I_0(k)\right|^2$}

  \includegraphics[width=8cm]{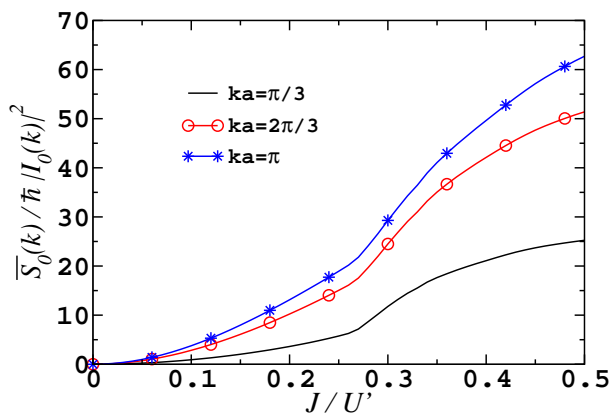}

\caption{(Color online) 
         Static structure factor as a function of $J$ in the hard-core limit at $T=0$
         averaged over $200$ disorder realizations.
         $L=200$, $L'=100$, $N=100$.
        }
\label{ssf}
\end{figure}

Finally, we consider the static structure factor defined as
\begin{equation}
S_0(k)=\int_{-\infty}^\infty S(k,E) d\,E
\;.
\end{equation}
In the case of hard-core bosons, it is given by Eq.~(\ref{dsfhc}), where
the $\delta$-function is formally replaced by $1$. 
Its $J$-dependence is shown in Fig.~\ref{ssf} for $ka=\pi/3, 2\pi/3$
and  $ka=\pi$. In general, it grows
monotonously with increasing $J$. Interestingly enough, we find a kink
which corresponds to the MI-BG transition point,  similar to the behavior
of $\bar P(0)$ in Fig.~\ref{md}.

\section{\label{Conclusions}Conclusions}

In the present work, we have studied quantum phase transitions of ultracold
bosons with repulsive interaction in a lattice with binary disorder.
The system is described by 
the Bose-Hubbard model with random on-site energies which follow a
binary probability distribution. The particular form of disorder
is physically realized, e.g.,
when two species of alkali-metal atoms with different masses are loaded
in an optical lattice. The latter is
created by counter-propagating laser beams which are
strongly  detuned from the atomic resonance.
If one of the species has a mass, say, $4$ times bigger
(as it is realized for the combination of $^{87}$Rb and $^{23}$Na atoms),
the tunneling of the heavier atoms in a deep enough lattice is  suppressed
by more than $3$ orders of magnitude compared to that of the lighter ones.
Thus, they effectively form immobile impurities interacting with the lighter atom species.
Larger mass differences, like for the combination of $^{40}$K-$^{7}$Li or
$^{87}$Rb-$^{7}$Li and $^{133}$Cs-$^7$Li,
render the difference in the tunneling rates even more drastic.
The impurity atoms then induce an effectively
quenched disorder potential for the lighter bosonic atoms.
When their number is less than the number of the lattice sites,
the probability to find more than one impurity at a lattice site can be extremely low.
This is garanteed by repulsive interactions between the bosonic impurities
and by Pauli's exclusion in the case of the fermionic ones.
Since the impurities are assumed immobile, their statistics does not play any role.
In fact, the interaction parameters $U$ and $U'$ can be tuned over a wider range
by the additional use of Feshbach resonances.

To calculate the boundaries of the MI phases in the phase diagram for arbitrary
lattice dimension $d$ at zero temperature, we have applied the
method of strong-coupling expansion.
We have shown that the MI phase exists also for incommensurate bosonic fillings,
and not only for  commensurate ones. Furthermore, the binary disorder
generates additional Mott lobes in the phase diagram.

For 1D lattices,
we have investigated the superfluidity of soft-core bosons at $T=0$
for the binary disorder.
Due to the exponential growth of the dimension of the bosonic Hilbert space
for  increasing boson numbers and numbers of lattice sites,
exact numerical diagonalization is possible only for small lattices,
which does not allow to control finite-size effects. However, 
the obtained results for small lattices 
are in good agreement with perturbative results obtained in  Sec.~\ref{SCE}
as well as with the exact results in the hard-core limit, see Sec.~\ref{BECSF}.

In the limit of infinitely strong repulsion (hard-core bosons), we have performed 
rather detailed exact studies by applying the Jordan-Wigner transformation. 
This allows to considerably  reduce
the computational complexity  since all the properties of
the strongly interacting system can be determined in terms of the solution
of the single-particle problem. The remaining disorder
average can straightforwardly be performed by standard numerical means. 
We have shown that the binary
disorder destroys the superfluidity in the thermodynamic limit in a similar manner as 
 for  Gaussian disorder studied earlier. However, in contrast to the case of   Gaussian
or uniform disorder,  we have found that the compressibility
of the BG phase can be extremely low. Several experimentally measurable quantities
such as the momentum distribution, the dynamical and static structure
factors and the density of excited states have been 
worked out. The MI-BG transition can be identified via rather sharp kinks in the
functional dependence of the maximum of the momentum distribution 
on the tunneling $J$ (Fig.~\ref{md}). Similar kinks occur in 
the static structure factor (Fig.~\ref{ssf}). These kinks allow to identify the 
MI-BG quantum phase transition. 
The energy-dependence of the dynamical structure factor yields complementary
information about the gap in the excitation spectrum in both phases (Fig.~\ref{dsf}).
Given the wide availability of elaborated experimental
techniques, we hope that
these predicted features will be found in real physical systems of ultracold atoms 
in the near future. 

In the present work, we did not consider the effects of harmonic confinement
which is normally present in most experiments with ultracold atoms in optical lattices.
One can expect that our results will remain valid for shallow traps.
On the other hand, coexistence of different phases in different spatial regions
can come into play as in the case without disorder~\cite{review,BDZ} depending on
the range of values of the local chemical potential in the region occupied by the atoms.
For instance, in the hard-core limit, there might be coexistence of the BG and MI with
one boson per site, if $J/U'>0.25$. In the opposite case $J/U'<0.25$, MI with non-commensurate
filling can coexist with the previous two phases. Detailed studies of the coexisting phases
and experimental signatures of the corresponding QPT is
a separate problem which requires further investigations.

\section*{Acknowledgment}

This work was supported by the SFB/TR 12 of the German Research Foundation (DFG).



\begin{thebibliography}{99}
\bibitem{note}
Similar treatment can be found in Ref.~\cite{MF} where the analysis
was restricted by $d=1$ and not all the possible values of the disorder
parameter $U'$ were explicitly considered.
\end{thebibliography}
\end{document}